\documentclass[submission,copyright,creativecommons]{eptcs}
\providecommand{\event}{PLACES 2020}

\usepackage{underscore} 
\usepackage{amssymb,amsmath}
\usepackage{listings}
\usepackage{url}
\usepackage[dvipsnames]{xcolor}
\usepackage{comment}
\usepackage{todonotes}
\usepackage{xspace}
\usepackage{cleveref} 
\usepackage{multicol}
\usepackage{wrapfig}
\usepackage{adjustbox} 
\usepackage{booktabs} 
\usepackage{makecell} 

\usepackage[T1]{fontenc}
\usepackage[scaled=.8]{beramono}

\newcommand{\ourlibrary}[0]{{\ttfamily SessionC\#}\xspace}
\newcommand{\myparagraph}[1]{\smallskip\noindent\textit{\bf #1\ }}
\newcommand{\etal}{{et al}.\@ }

\makeatletter

\newenvironment{btHighlight}[1][]
{\begingroup\tikzset{bt@Highlight@par/.style={#1}}\begin{lrbox}{\@tempboxa}}
{\end{lrbox}\bt@HL@box[bt@Highlight@par]{\@tempboxa}\endgroup}

\newcommand\btHL[1][]{%
  \begin{btHighlight}[#1]\bgroup\aftergroup\bt@HL@endenv%
}

\def\bt@HL@endenv{%
  \end{btHighlight}%
  \egroup
}
\newcommand{\bt@HL@box}[2][]{%
  \tikz[#1]{%
    \pgfpathrectangle{\pgfpoint{1pt}{0pt}}{\pgfpoint{\wd #2}{\ht #2}}%
    \pgfusepath{use as bounding box}%
    \node[anchor=base west, fill=orange!30,outer sep=0pt,inner xsep=1pt, inner ysep=0pt, rounded corners=3pt, minimum height=\ht\strutbox+1pt,#1]{\raisebox{1pt}{\strut}\strut\usebox{#2}};
  }%
}

\makeatother

\lstdefinestyle{sharpc}{
  language=[Sharp]C,
  basicstyle=\ttfamily\footnotesize,
  numbers=left,
  numberstyle=\ttfamily\scriptsize,
  morekeywords={var,async,await,Task},
  morekeywords=[2]{Send,Recv,Receive,ReceiveAsync,Deleg,DelegRecv,DelegRecvAsync,
    DelegNew,Select,SelectLeft,SelectRight,Offer,OfferAsync,End,
    ForkThread,Goto0,Goto1,Goto2,Goto3,Close,Dual,Eps
  },
  morekeywords=[3]{left, right, deleg, chan},
  keywordstyle=\color{magenta},
  keywordstyle=[2]\color{blue},
  keywordstyle=[3]\color{OliveGreen}\textit,
  moredelim=[is][\color{red}\bfseries]{@@}{@@},
  moredelim=[is][\color{red}]{--}{--},
  moredelim=[is][\itshape]{__}{__},
  moredelim=**[is][\btHL]{`}{`},
  commentstyle=\color{OliveGreen},
  escapeinside={^}{^},
  mathescape=true,
  showspaces=false,
  showstringspaces=false,
  xleftmargin=14pt,
  numbersep=6pt,
}
\title{Fluent Session Programming in C\#}
\author{
Shunsuke Kimura \qquad\qquad Keigo Imai
\institute{Gifu University, Japan}
\email{\quad kimura@ct.info.gifu-u.ac.jp \quad\qquad keigoi@gifu-u.ac.jp}
}
\def\titlerunning{Fluent Session Programming in C\#}
\def\authorrunning{S. Kimura \& K. Imai}

\begin{document}

\maketitle

\begin{abstract}
We propose \ourlibrary, a lightweight session typed library for safe
concurrent/distributed programming.  The key features are (1) the
improved fluent interface which enables writing communication in
chained method calls, by exploiting C\#'s out variables, and (2)
amalgamation of session delegation with async/await, which
materialises session cancellation in a limited form,
which we call {\em session intervention}. We
show the effectiveness of our proposal via a Bitcoin miner
application.
\end{abstract}

\section{Introduction}\label{sec:introduction}

Session types \cite{honda98} are a theoretical framework for statically specifying and verifying communication protocols
in concurrent and distributed programs.
Session types guarantee that a well-typed program follows a {\em safe} communication protocol free from reception errors (unexpected messages) and deadlocks.

The major gaps between session types and ``mainstream'' programming language type system are the absence of the two key features:
(1) {\em duality} for checking the communication protocol realises reciprocal communication actions between two peers,
and
(2) {\em linearity}
ensuring that each peer is exactly following the protocol,
in the way that channel variables are exclusively used from one site for the exact number of times.
Various challenges have been made for incorporating them into general-purpose languages
including Java \cite{sessionjava}, Scala \cite{scalas16lightweight},
Haskell \cite{haskell08,imai10session,lindley16embedding,orchard16effects},
OCaml \cite{padovani16simple,imai18sessionscp} and Rust \cite{jespersen15session,DBLP:journals/corr/abs-1909-05970}.

We observe that the above-mentioned gaps in session-based programming can be
{\em narrowed} further by the recent advancement of programming languages,
which is driven by various real-world programming issues.
In particular, C\#~\cite{csharp} is widely used
in areas ranging from Windows and web application platforms to gaming (e.g. Unity),
and known to be eagerly adopting various language features
including
async/await,
reifiable generics, named parameters, \lstinline!out! variables and extension methods.


In this paper, we propose {\bfseries\ourlibrary} -- a library implementation of session types on top of the rich set of features in C\#,
and show its usefulness in concurrent/distributed programming,
aiming for {\em practicality}.
\begin{wrapfigure}{r}{0.22\textwidth}
  \vspace*{-0.6cm} 
  \begin{center}
    \includegraphics[trim=1.8cm 0.6cm 1.2cm 1.2cm,width=3.5cm,clip]{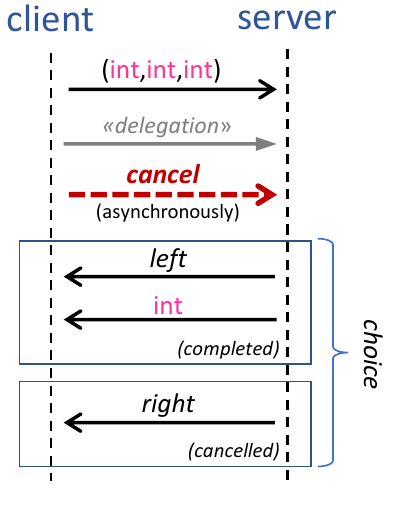}
    \vspace*{-1.3cm} 
  \end{center}
\end{wrapfigure}
Namely, (1)
it has an improved {\em fluent interface} (i.e., method calls can be chained) 
via C\#'s \lstinline!out! variables,
reducing the risk of linearity violation in an {\em idiomatic} way.
Furthermore, (2) it enables {\em session cancellation} in a limited form
--- which we call {\em session intervention} ---
by utilising
amalgamation of C\#'s async/await and {\em session delegation} in thread-based concurrency.

We illustrate the essential bits of \ourlibrary where
a {\em cancellable} computation is
guided by
session types, by a use-case where
a C\# thread calculates a cancellable {\tt tak} function
which is designed to have a long running time \cite{tarai}.
The figure on the right
depicts the overall communication protocol,
which can be written in \ourlibrary as a \textbf{\em protocol specification}
describing  a client-server communication protocol
from the client's viewpoint,
as follows:
\begin{lstlisting}[numbers=none,xleftmargin=8pt]
      var prot = Send(Val<(int,int,int)>, Deleg(chan:Recv(Unit,End),
                                            Offer(left:Recv(Val<int>, End), right:End)));
\end{lstlisting}
From the above, C\# compiler can {\em statically} derive both the client and the server's {\em session type}
which is {\em dual} to each other, ensuring the safe interaction between the two peers.
The client starts with an output
(\lstinline!Send!)
of a triple of integer values (\lstinline!Val<(int,int,int)>!)
as arguments to {\tt tak} function,
which continues to a {\em session delegation} (\lstinline!Deleg!)
where a channel with an input capability (\lstinline!Recv(Unit,End)!) is passed --- annotated by a {\em named parameter} \lstinline!chan! --- from the client to the server
so that the server can get notified of {\em cancellation}.
\lstinline!Offer! specifies the client offering a binary choice between
\lstinline!left! option with
a reception
(\lstinline!Recv!)
of the resulting \lstinline!int! value,
and \lstinline!right! option with an immediate closing
(\lstinline!End!),
in case {\tt tak} is cancelled.

\begin{figure}[tb]
 \begin{lstlisting}[multicols=2]
var cliCh = prot.ForkThread(srvCh => {^\label{line:tarai:impl:start}^^\label{line:tarai:anon:start}^
 var srvCh2 = ^\label{line:tarai:srv:first}^
  srvCh.Receive(out int x, out int y, out int z)^\label{line:tarai:srv:receive}^
       .DelegRecv(out var --cancelCh--);^\label{line:tarai:srv:delegnew}^
 --cancelCh--.ReceiveAsync(out Task @@cancel@@).Close();^\label{line:tarai:srv:cancelCh}^
 try
 {
  var result = Tak(x, y, z);  // compute tak ^\label{line:tarai:srv:takcall}^
  srvCh2.SelectLeft().Send(result).Close();^\label{line:tarai:srv:success}^
 }
 catch (OperationCanceledException)^\label{line:tarai:srv:catch:start}^
 {
  srvCh2.SelectRight().Close(); // if cancelled^\label{line:tarai:srv:cancel}^
 }^\label{line:tarai:srv:catch:end}^
 int Tak(int a, int b, int c) {^\label{line:tarai:srv:takbegin}^
  if (@@cancel@@.IsCompleted)^\label{line:tarai:srv:iscompleted}^
   throw new OperationCanceledException();^\label{line:tarai:srv:exception}^
  return a <= b ? b :^\label{line:tarai:srv:result1}^
   Tak(Tak(a-1,b,c),Tak(b-1,c,a),Tak(c-1,a,b));^\label{line:tarai:srv:result2}^
 }});^\label{line:tarai:srv:takend}^^\label{line:tarai:anon:end}^^\label{line:tarai:impl:end}^
\end{lstlisting}
\caption{A Cancellable {\tt tak} \cite{tarai} Implementation in \ourlibrary\label{fig:tarai}}
\end{figure}

The {\tt tak} server's \textbf{\em endpoint implementation} in \Cref{fig:tarai}
enjoys {\em compliance} to the protocol {\tt prot} above,
by starting itself using \lstinline!ForkThread! method of protocol \lstinline!prot!.
It runs an anonymous function communicating on a
channel \lstinline!srvCh! passed as an object
with the exact communication API methods
prescribed by the session type, which is derived from the protocol specification.
Note that the channel \lstinline!cliCh! returned by \lstinline!ForkThread!
has the session type enforcing the client as well (which will be shown by \Cref{fig:taraiclient} of \S~\ref{subsubsec:takclient}).


Notably,
the use of \textbf{\em improved fluent interface}
in Line~\ref{line:tarai:srv:receive}-\ref{line:tarai:srv:delegnew}
enhances protocol compliance,
where the consecutive {\em input} actions (\lstinline!Receive!) are realised as the {\em chained} method calls in a row,
promoting {\em linear} use of the returned session channels.
The \lstinline!out! keywords in Line~\ref{line:tarai:srv:receive} are the key for this;
they declare
the three variables \lstinline!x!, \lstinline!y! and \lstinline!z!
{\em in-place},
and upon delivery of the integer triple,
the received values are bound to these variables (as their references are passed to the callee).
In Line~\ref{line:tarai:srv:delegnew},
\lstinline!DelegRecv! accepts a delegation from the client, binding it to \lstinline!--cancelCh--!.
The protocol for \lstinline!--cancelCh--! is inferred via \lstinline!var! keyword
as \lstinline!Recv(Unit,End)! specifying the reception of a cancellation.
The continuation is then assigned to \lstinline!srvCh2!.

\begin{wrapfigure}{r}{0.24\textwidth}
  \vspace*{-0.5em} 
  \includegraphics[scale=0.5]{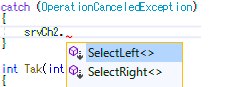}
  \vspace*{-2em} 
  \caption{Completion\label{fig:completion}}
  \vspace*{-1em} 
\end{wrapfigure}
In addition, our design of the fluent interface also takes advantage of the modern {\em programming environment}
like Visual Studio, via \textbf{\em code completion}:
The code editor suggests the correct communication primitive to the
programmer, guided by the session types.
The screenshot in \Cref{fig:completion} is such an example in Visual Studio
where the two alternatives in a \lstinline!Select! branch are suggested
right after the symbol `\lstinline!.!' (dot) is typed.

Furthermore, we claim that the \textbf{\em session intervention pattern}
emerging from Line~\ref{line:tarai:srv:cancelCh}
is a novel intermix of C\#'s async/await and session types in \ourlibrary,
where a control flow of a session is
affected by a delegated session.
The delegated session can be seen as a {\em cancellation token} in folks, modelled by session-types.
Line~\ref{line:tarai:srv:cancelCh}
schedules {\em asynchronous} reception of cancellation on
\lstinline!--cancelCh--!
(\lstinline!ReceiveAsync!),
immediately returning from the method call (i.e., non-blocking)
and binding the variable \lstinline!@@cancel@@! to a \lstinline!Task!.
The task \lstinline!@@cancel@@! gets {\em completed}
when a \lstinline!Unit! value of cancellation is delivered.
The following \lstinline!Close! leaves the reception incomplete.
The task is checked inside \lstinline!Tak! function (Line~\ref{line:tarai:srv:iscompleted}),
raising \lstinline!OperationCanceledException! if cancellation is delivered before finishing the calculation,
which is caught by the outer \lstinline!catch! block in Lines~\ref{line:tarai:srv:catch:start}-\ref{line:tarai:srv:catch:end}.
Lines~\ref{line:tarai:srv:takcall}-\ref{line:tarai:srv:success} in a \lstinline!try! block
calls the \lstinline!Tak! function and sends (\lstinline!Send!) the result back to the client after selecting the left option (\lstinline!SelectLeft!).
If a cancellation is signalled by an exception,
in \lstinline!catch! block (Line~\ref{line:tarai:srv:cancel})
the server selects the right option (\lstinline!SelectRight!) and closes the session.

See that
all interactions in \Cref{fig:tarai} are deadlock-free,
albeit binary session type systems like \cite{honda98} and its successors
used in many
libraries, including ours, do not prevent deadlocks in general\footnote{
  Exceptions are GV \cite{DBLP:journals/jfp/Wadler14,DBLP:conf/esop/LindleyM15} and its successors (e.g. EGV \cite{fowler}),
  and Links language \cite{DBLP:conf/fmco/CooperLWY06}.
}.
In the case above, operations on the delegated session \lstinline!--cancelCh--! are non-blocking,
and the session type on \lstinline!srvCh! guarantees progress,
provided that the client respects the {\em dual} session types as well,
which is the case in the client in \Cref{fig:taraiclient}
shown later (\S~\ref{subsubsec:takclient}).
Note that, however, in general, the program using {\em blocking} operations of \lstinline!Task! may cause a deadlock.

\myparagraph{Notes on Nondeterminism.}
The session intervention above is {\em nondeterministic}, as the client can send a cancellation {\em at any time}
after it receives \lstinline!--cancelCh--!.
There is a possibility where the server may {\em disregard} the cancellation.
For example, the server will {\em not} cancel the calculation if the client outputs
on \lstinline!--cancelCh--! {\em after} the server check \lstinline!--cancel--! on Line~\ref{line:tarai:srv:iscompleted},
and after the result has been computed (i.e., no recursive call is made at Lines~\ref{line:tarai:srv:result1}-\ref{line:tarai:srv:result2}).
In this case, the cancellation is still delivered to the server, and silently {\em ignored}.
Note that the channel \lstinline!--cancelCh--! is still used faithfully according to its protocol \lstinline!Recv(Unit,End)! and
{\em session fidelity} is maintained, as the reception is already ``scheduled'' by \lstinline!ReceiveAsync!
on Line~\ref{line:tarai:srv:cancelCh}.
Note also that there is {\em no} confusion that the client consider that the calculation is cancelled,
as the client must check the result of a cancellation via \lstinline!Offer!,
which we will revisit in \S~\ref{subsubsec:takclient}.

\myparagraph{Notes on Linearity.}
C\# does not have linear types, as in the most of mainstream languages.
Thus there are two risks of linearity violations which are not checked statically:
(1) use of a channel more than once and (2) channels discarded without using.
For (1), we have implemented dynamic checking around it,
which raises \lstinline!LinearityViolationException! when a channel is used more than once.
Regarding (2), although current \ourlibrary does not have capability to check it,
we are planning to implement it around the {\em destructor} of a channel
which is still not optimal, however better than nothing, as the check is delayed to the point when the garbage collector is invoked.

\myparagraph{Notes on Session Cancellation.}
There are a few literature on session cancellation, such as Fowler et al.'s EGV \cite{fowler}, which we do not follow for now.
Instead, The session intervention pattern above uses branching (\lstinline!Select!/\lstinline!Offer!) to handle a cancellation.
There are a few issues on session cancellation in this form:
(a) The cancellation handler clutters the protocol as the number of interaction increases, as mentioned in \cite[\S~1.2]{fowler}.
Although the branching-based solution is suitable for a short protocol like the above,
there is a criticism by a reviewer specifically to \ourlibrary that
(b) it lacks an exception handling mechanism, such as {\em crashing} (e.g. unhandled exceptions) and {\em disconnecting} (e.g. TCP connection failure).
While we are yet to implement exception handling mechanisms, the distributed version of \ourlibrary equips \lstinline!SessionCanceller!
which handles session disconnection in terms of \cite{fowler,DBLP:journals/corr/abs-1909-05970}.

Based on the key features and notes above, in the following sections,
we explore the design space of
modern session programming in \ourlibrary,
showing the effectiveness of our proposal.
The rest of this paper is structured as follows:
In \S~\ref{sec:basic}, we describe the basic design of \ourlibrary,
and show a few application in \S~\ref{sec:application}.
We conclude with remarks in \S~\ref{sec:conclusion}.
Appendix \S~\ref{app:combinators} describes implementation detail of the protocol combinators,
and Appendix \S~\ref{app:delegation} discusses more details on delegating a recursive session.
Appendix \S~\ref{app:moreexamples} includes more examples using \ourlibrary, including
distributed implementation.
The \ourlibrary is available at the following URL:
\begin{center}
  \url{https://github.com/curegit/session-csharp/}
\end{center}

\section{Design of \ourlibrary}\label{sec:basic}
In this section, we show the design of \ourlibrary
which closely follows Honda et al.'s binary session types \cite{honda98}.
\S~\ref{subsec:protocolcombinators} introduces protocol combinators,
by following Pucella and Tov's approach \cite[\S~5.2]{haskell08}
with a few extensions including recursion and delegation.
\S~\ref{subsec:fluentapi} introduces the improved fluent API,
taking inspiration from Scribble \cite{scribble,hy16}
and process calculi's literature.
\S~\ref{subsec:recursion} discusses
an encoding of mutually recursive sessions with less notational overhead.

\subsection{Session Types, Protocol Combinators and Duality}
\label{subsec:protocolcombinators}

\begin{figure}[t]
  \begin{center}
  \begin{adjustbox}{width=\columnwidth,center}
    \begin{tabular}{ll|ll}
      Session type & Synopsis & Combinator$^*$ & Return type$^*$\\
      \toprule
      \lstinline!Send<V,S>! & Output \lstinline!V! then do \lstinline!S!
      & \lstinline!Send(v,p)$^{(1),(2)}$!
      & \lstinline!Dual<Send<V,S>,Recv<V,T>>!\\
      \lstinline!Recv<V,S>! & Input \lstinline!V! then do \lstinline!S!
      & \lstinline!Recv(v,p)$^{(1),(2)}$!
      & \lstinline!Dual<Recv<V,S>,Send<V,T>>!\\
      \hline
      \lstinline!Select<SL,SR>! & \makecell[l]{Internal choice\\ between \lstinline!SL! and \lstinline!SR!}
      & \lstinline!Select(left:p$_L$,right:p$_R$)$^{(3)}$!
      & \lstinline!Dual<Select<S$_L$,T$_L$>,Offer<S$_R$,T$_R$>>!\\
      \lstinline!Offer<SL,SR>! & \makecell[l]{External choice\\ between \lstinline!SR! and \lstinline!SR!}
      & \lstinline!Offer(left:p$_L$,right:p$_R$)$^{(3)}$!
      & \lstinline!Dual<Offer<S$_L$,T$_L$>,Select<S$_R$,T$_R$>>!\\
      \hline
      \lstinline!Eps! & End of the session
      & \lstinline!End!
      & \lstinline!Dual<Eps,Eps>!\\
      \lstinline!Goto0! & Jump to the beginning
      & \lstinline!Goto0!
      & \lstinline!Dual<Goto0,Goto0>!\\
      \hline
      \lstinline!Deleg<S0,T0,S>!
      & \makecell[l]{Delegate \lstinline!S0! then do \lstinline!S!\\(where \lstinline!T0! is dual of \lstinline!S0!)}
      & \lstinline!Deleg(chan:p$_0$,p)$^{(2),(3)}$!
      & \lstinline!Dual<Deleg<S$_0$,T$_0$,S>,DelegRecv<S$_0$,T>>!\\
      \lstinline!DelegRecv<S0,S>!
      & \makecell[l]{Accept delegation \lstinline!S0!\\ then do \lstinline!S!}
      & \lstinline!DelegRecv(chan:p$_0$,p)$^{(2),(3)}$!
      & \lstinline!Dual<DelegRecv<S$_0$,S>,Deleg<S$_0$,T$_0$,T>>!\\
    \end{tabular}
  \end{adjustbox}
  \end{center}
  \small{
    *Note: The right half of the table assume that
    (1) variable \lstinline!v! has type \lstinline!Val<V>!,
    (2) variable \lstinline!p! has type \lstinline!Dual<S,T>!,
    (3) variable \lstinline!p$_i$! has type \lstinline!Dual<S$_i$,T$_i$>! for $i\in\{L,R,0\}$.
  }
  \caption{Session Types and Protocol Combinators\label{fig:sessiontypes}}
\end{figure}

Duality is the key to ensure that a pair of session types realise a safe series of interaction.
Before introducing protocol combinators, we summarise session types in \ourlibrary in the left
half of \Cref{fig:sessiontypes}.  Type \lstinline!Send<V,S>! and
\lstinline!Recv<V,S>! are output and input of
value of type \lstinline!V!, respectively, which continues
to behave according to the session type \lstinline!S!.
\lstinline!Select<SL,SR>!
means that a process internally {\em decides} whether to behave according to \lstinline!SL! or \lstinline!SR!,
by sending either label of \lstinline!left! or \lstinline!right!, which is called as {\em internal choice}.
\lstinline!Offer<SL,SR>! is an {\em external choice} where a process
offers to its counterpart two possible behaviours \lstinline!SL! and \lstinline!SR!.
\lstinline!Eps! is the end of a session.
\lstinline!Goto0! specifies transition to the beginning of the session,
which makes a limited form of {\em recursive session}.
Later on, we extend this to mutual recursion by having more than one session types in a C\# type
and accessing them via an index,
which is why we annotate \lstinline!0! as the suffix to \lstinline!Goto!.
\lstinline!Deleg<S0,T0,S>! is a {\em delegation} of session \lstinline!S0! which continues to \lstinline!S!,
where the additional parameter \lstinline!T0! is the dual of \lstinline!S0!.

Note that it is possible to implement delegation {\em without} \lstinline!Deleg! and \lstinline!DelegRecv!, but with \lstinline!Send! and \lstinline!Recv! instead.
The sole purpose of this distinction is the parameter \lstinline!T0!, which is used by \lstinline!DelegNew!,
which we will develop later, to give the dual session type to the freshly created channel without further protocol annotation.
\lstinline!DelegRecv<S0,S>! is an acceptance of delegation of session \lstinline!S0! which continues to \lstinline!S!.

We illustrate our protocol combinators in the right half of \Cref{fig:sessiontypes},
making them {\em prove} duality of two types
by restricting the constructors of  \lstinline!Dual<S,T>! to them
having \lstinline!S! and \lstinline!T! to be dual to each other.
In the ``Combinator'' column of \Cref{fig:sessiontypes},
the intuitive meaning of each protocol combinator can be understood as the session type in the same row of the left half
specifying the {\em client side}'s behaviour.
The ``Return type'' column establishes duality,
by pairing each session type in the first parameter for the client with the reciprocal behaviours in the second one for the server.
The type \lstinline!Val<V>! is the placeholder for payload types of \lstinline!Send! and \lstinline!Recv!.
For example, \lstinline!Send(v,p)! with type \lstinline!Dual<Send<V,S>,Recv<V,T>>! establishes the duality between
two session types \lstinline!Send<V,S>! and \lstinline!Recv<V,T>! provided that
\lstinline!S! and \lstinline!T! are dual to each other,
which is ensured by the nested protocol object \lstinline!p!.
We defer the actual method signatures of protocol combinators to Appendix \S~\ref{app:combinators}.

\subsection{A Fluent Communication API}
\label{subsec:fluentapi}

\begin{figure}
  \begin{center}
    \begin{tabular}{l|l}
      \multicolumn{2}{l}{\textbf{Creating a session}}\\
      \hline
      \multicolumn{2}{l}{\lstinline!var cliCh = prot.ForkThread(srvCh => {__stmts__})!}\\
      \midrule
      \multicolumn{2}{l}{\textbf{Communication}}\\
      \hline
      Session type & Method\\
      \hline
      \lstinline!Send<V,S>! & \lstinline!ch.Send(v)!\\
      \lstinline!Recv<V,S>! & \lstinline!ch.Receive(out V x)!, \lstinline!ch.ReceiveAsync(out Task<V> task)!\\
      \hline
      \lstinline!Eps! & \lstinline!ch.Close()!\\
      \lstinline!Goto0! & \lstinline!ch.Goto0()!\\
      \hline
      \lstinline!Select<SL,SR>! & \lstinline!ch.SelectLeft()!, \lstinline!ch.SelectRight()!\\
      \lstinline!Offer<SL,SR>! & \lstinline!ch.Offer(left:__leftFunc__, right:__rightFunc__)!\\
      \lstinline!Deleg<S0,T0,S>! & \lstinline!ch.Deleg(ch2)!, \lstinline!ch.DelegNew(out Session<T0,T0> ch2)!\\
      \lstinline!DelegRecv<S0,S>! & \lstinline!ch.DelegRecv(out Session<S0,S0> ch2)!
    \end{tabular}
    \caption{The Communication API of \ourlibrary\label{fig:api}}
  \end{center}
\end{figure}

In \Cref{fig:api}, we show the communication API of \ourlibrary which we develop in this subsection.
The first column of the figure specifies the session type of the method in the second column.
The fluent interface contributes to reducing the risk of linearity violation,
by returning the channel with a continuation session type
which increase the opportunity to chain the method call.
An exception is \lstinline!Offer!
which takes two functions \lstinline!__leftFunc__! and \lstinline!__rightFunc!
taking a channel with different continuation session type
for selection labels \lstinline!left! and \lstinline!right!, respectively.



\subsubsection{Channels and Threads Maintaining Duality}
The {\em channel type} \lstinline!Session<S,E>! plays
the key role in maintaining a session's evolution in the recursive type structure,
where the type parameter \lstinline!S! is the session type assigned to the channel,
while \lstinline!E! is the {\em session environment} of a channel
which serves as a table for recursive calls (\lstinline!Goto!) to look up the {\em next} behaviour.
In other words, the \lstinline!S!-part progresses
when the interaction occurs on that channel,
while the \lstinline!E!-part {\em persists} (i.e., remains unchanged) during a session,
maintaining the global view of a session.
Thus, for example, in a method call \lstinline!ch.Send(v)!,
the channel \lstinline!ch! must have type  \lstinline!Session<Send<V,S>,E>!,
which returns \lstinline!Session<S,E>!.
We explain how the recursive structure is maintained later in \S~\ref{subsec:recursion}.

Based on the duality established by the protocol combinators,
the \lstinline!ForkThread! method ensures
{\em safe} communication on \lstinline!Session<S,E>! channels between the main thread and the forked threads.
Concretely, provided \lstinline!prot! has \lstinline!Dual<S,T>!
saying \lstinline!S! and \lstinline!T! are dual to each other,
a method call \lstinline!prot.ForkThread(ch => $\textit{stmt}$)!
forks a new server thread, running \lstinline!$\textit{stmt}$!
with channel \lstinline!ch! of type \lstinline!Session<T,T>!,
returning the other end of channel \lstinline!Session<S,S>!.
The \lstinline!ForkThread! is defined in the following way:
\begin{lstlisting}[numbers=none,xleftmargin=8pt]
class Dual<S,T> {  static Session<S,S> ForkThread(Func<Session<T,T>> fun) { ... }  }
\end{lstlisting}
Note that the part \lstinline!<S,S>! (and \lstinline!<T,T>!) requires
the beginning of a session being the same as in the session environment,
maintaining the recursive structure by
specifying \lstinline!Goto0! going back to \lstinline!S!
(and \lstinline!T! resp.).

\subsubsection{Protocol Compliance via Extension Methods}

The communication API enforces {\em compliance} to the type parameters in \lstinline!Session<S,E>!,
via {\em extension methods} of that type
which can have additional constraints on type parameters.
An extension method is the one which can be added to the existing class without modifying the existing code.
For example, the following method declaration adds a method to
\lstinline!List<T>! class in the standard library:
\begin{lstlisting}[numbers=none,xleftmargin=8pt]
static void AddInt(this List<int> intList, int x) { ... }
\end{lstlisting}
The \lstinline!this! keyword in the first parameter specifies the method as an extension method,
where the possible type of \lstinline!obj! is restricted to \lstinline!List<int>!.
In this way, we declare the fluent API of output \lstinline!ch.Send(v)!, for example, as follows:
\begin{lstlisting}[numbers=none,xleftmargin=8pt]
static Session<S,E> Send<V,S,E>(this Session<Send<V,S>,E> ch, V v) { ... }
\end{lstlisting}


\subsubsection{Binders as \lstinline!out! Parameters, and Async/Await Integration}\label{subsubsec:integration}
One of the central ideas of the fluent API in \ourlibrary is to exploit C\#'s
\lstinline!out! method parameter to increase chances for method chaining.
This is mainly inspired by Scribble \cite{scribble,hy16} implemented in Java,
however,
thanks to the \lstinline!out! parameter in C\#,
there is no need to explicitly passing a {\em buffer} to receive
an input value as in Java,
keeping the session-typed program more concise and readable.
\lstinline!Receive! and the acceptance of delegation \lstinline!DelegRecv! are implemented similarly,
in the following way:
\begin{lstlisting}[numbers=none,xleftmargin=8pt]
static Session<S,E> Receive<V,S,E>(this Session<Recv<V,S>,E> ch, out V v) { ... }
static Session<S,E> DelegRecv<S0,S,E>(this Session<DelegRecv<S0,S>,E> ch,
                                      out Session<S0,S0> ch2) { ... }
\end{lstlisting}

More interestingly,
the \lstinline!out! parameter in the method call \lstinline!obj.Receive(out var x)! resembles {\em binders} in process calculi,
like an input prefix $a(x).P$ in the $\pi$-calculus.
By expanding this observation to
name restriction $(\nu x)P$ in the $\pi$-calculus and other constructs in literature,
we crystallise a few useful communication patterns of
process calculi in \ourlibrary; namely
(1) {\em bound output} and (2) {\em delayed input},
where the latter is implemented using async/await.

Bound output is a form of channel-passing
where the freshly-created channel is passed immediately
through another channel, which is written in the $\pi$-calculus as
$(\nu x)\overline{a}x.P$,
and $\overline{a}(x).P$ in short.
As it leaves the {\em other end} of a channel at the sender's side, we need the {\em dual}
of the carried (delegated) session type, which is why we have
both carried type \lstinline!S0! and its dual \lstinline!T0! in a delegation type \lstinline!Deleg<S0,T0,S>!.
Thus, delegation \lstinline!Deleg! and its bound-output variant \lstinline!DelegNew! is defined as follows:
\begin{lstlisting}[numbers=none,xleftmargin=8pt]
static Session<S,E> Deleg<S0,T0,S,E>(this Session<Deleg<S0,T0,S>,E> ch, Session<S0,S0> ch2) { ... }
static Session<S,E> DelegNew<S0,T0,S,E>(this Session<Deleg<S0,T0,S>,E> ch,
                                        out Session<T0,T0> ch2) { ... }
\end{lstlisting}
See that \lstinline!DelegNew! declares the \lstinline!out! parameter in the second one,
where it binds the dual type \lstinline!T0! of the delegated type \lstinline!S0!.


The \lstinline!ReceiveAsync!
is a form of delayed input in the $\pi$-calculus literature \cite[\S~9.3]{merro-sangiorgi-2004},
also inspired by
Scribble's {\em future} \cite[\S~13.4]{BETTYTOOLBOOK:scribble}. \nocite{BETTYTOOLBOOK}
The delayed input
{\em asynchronously} inputs a value i.e.,
the execution progresses without waiting for delivery of an input value,
which blocks at the place it uses the input variable.
This is realised by method call \lstinline!ch.ReceiveAsync(out Task<V> task)! which
binds a fresh {\em task} to variable \lstinline!task! which completes when the value is delivered.
We illustrate the signature of \lstinline!ReceiveAsync! in the following\footnote{
  \Cref{fig:tarai} uses the overloaded version where payload type \lstinline!V! is fixed to \lstinline!Unit!,
  having \lstinline!Task! instead of \lstinline!Task<V>! in the second argument.
}:
\begin{lstlisting}[numbers=none,xleftmargin=8pt]
static Session<S,E> ReceiveAsync<V,S,E>(this Session<Recv<V,S>,E> ch, out Task<V> v) { ... }
\end{lstlisting}
Note that the implementation adheres the communication pattern specified in a session type,
as the subsequent communication on the same channel
does not take place until the preceding reception occurs.


\subsubsection{A {\ttfamily tak} Client Example}\label{subsubsec:takclient}
Based on the communication API shown in this section, including \lstinline!Offer! and \lstinline!DelegNew!,
we show an implementation of {\tt tak} client in \Cref{fig:taraiclient},
with a timeout.
Line~\ref{line:takcli:send} sends the three arguments \lstinline!(16,3,2)! to the server, and
Line~\ref{line:takcli:delegnew} freshly creates a channel \lstinline!cancelCh! and send it to the server using bound output \lstinline!DelegNew!,
for later termination request.
Lines~\ref{line:takcli:delay}-\ref{line:takcli:stop} arranges an output of a termination request in 10 seconds
(\lstinline!10000! milliseconds).
\lstinline!Offer! on Line~\ref{line:takcli:offer} makes an external choice on a channel.
The \lstinline!left! case on Lines~\ref{line:takcli:leftstart}-\ref{line:takcli:leftend}
handles the successful completion of the calculation, where the client
receives the result \lstinline!ans! and print it on the screen.
The \lstinline!right! case (Lines~\ref{line:takcli:rightstart}-\ref{line:takcli:rightend})
immediately closes the channel and prints \lstinline!"Cancelled"! on the console.

As we also noted in Introduction, the cancellation request may be disregarded by the server
if she has already finished the calculation.
Also note that the delegated channel \lstinline!cancelCh! must be used according to the linearity constraint (of which dynamic checking in \ourlibrary is yet to be implemented though),
even if the client does not wish to cancel the calculation.
In that case, the client can send a dummy cancellation request {\em after} it receives the result.

\begin{figure}
  \begin{lstlisting}[multicols=2]
var cliCh2 = cliCh.Send((16, 3, 2))^\label{line:takcli:send}^
                  .DelegNew(out var cancelCh);^\label{line:takcli:delegnew}^
Task.Delay(10000).ContinueWith(_ => {^\label{line:takcli:delay}^
  cancelCh.Send().Close(); });^\label{line:takcli:stop}^
cliCh2.Offer(^\label{line:takcli:offer}^
  left: cliCh3 => {^\label{line:takcli:leftstart}^
    cliCh3.Receive(out var ans).Close();
    Console.WriteLine("Tak(16,3,2) = " + ans);^\label{line:takcli:leftend}^
}, right: cliCh3 => {^\label{line:takcli:rightstart}^
    cliCh3.Close();
    Console.WriteLine("Cancelled");
});^\label{line:takcli:rightend}^
  \end{lstlisting}
  \caption{A {\tt tak} Client with a Timeout\label{fig:taraiclient}}
\end{figure}

\subsection{Recursive Sessions, Flatly}\label{subsec:recursion}
To handle mutually-recursive structure of a session,
we extend the session environment to have more than one session type.
We extend the notion of duality to the tuple of session types,
and provide the protocol combinator \lstinline!Arrange(p1,p2,..)!,
where \lstinline!p1!, \lstinline!p2!, $\ldots$ refers to them each other via
\lstinline!Goto1!, \lstinline!Goto2!, $\ldots$.
For example, a protocol specification which alternately sends and receives an integer is written as follows:
\begin{lstlisting}[numbers=none,xleftmargin=8pt]
var prot = Arrange(Send(Val<int>, Goto2), Recv(Val<int>, Goto1));
\end{lstlisting}
Note that the indices origin from one,
to avoid confusion in a session environment with the single-cycled sessions
using \lstinline!Goto0!.

The main difference from the one by Pucella and Tov \cite{haskell08}
is that, to avoid notational overhead, we stick on {\em flat} tuple-based representation \lstinline!(S0,S1,..)!
rather than a nested cons-based list \lstinline!Cons<S0,Cons<S1,..>>!.
This also elides {\em manual} unfolding of a recursive type from $\mu\!\alpha.T$
to $T[\mu\!\alpha.T/\alpha]$ encoded as \lstinline!enter! in \cite{haskell08},
resulting in a less notational overhead in recursive session types than \cite{haskell08}.
This ad-hoc encoding comes at a cost;
the number of cycles in a recursive session is limited because the size of the tuple is limited,
we must overload methods since we do not have a structural way to manipulate tuple types
-- although the maximum size of tuples of 8-9 seems enough
for a tractable communication program.
Keeping this in mind,
the duality proof, \lstinline!DualEnv<SS,TT>!,
which states the duality between the tuple of session types \lstinline!SS! and \lstinline!TT!,
as well as \lstinline!ForkThread! and \lstinline!Goto! methods for mutually recursive sessions
are implemented as follows:
\begin{lstlisting}[numbers=none,xleftmargin=8pt]
static DualEnv<(S1,S2),(T1,T2)> Arrange<S1,S2,T1,T2>(Dual<S1,T1> p1, Dual<S2,T2> p2) { ... } ...
static Session<S1,(S1,S2)> ForkThread<S1,S2,T1,T2>(this DualEnv<(S1,S2),(T1,T2)> prot,
                                                   Func<Session<T1,(T1,T2)>> fun) { ... }
static Session<S1,(S1,S2)> Goto1<S1,S2>(this Session<Goto1,(S1,S2)> ch) { ... }
static Session<S2,(S1,S2)> Goto2<S1,S2>(this Session<Goto2,(S1,S2)> ch) { ... }
\end{lstlisting}
The overloaded versions up to 8-ary is defined in similar way.

\myparagraph{Notes on Structural Recursion in C\#.}
A reviewer mentioned that there should be an encoding using recursive generic types in C\#.
For example, it would be possible to declare the following session type in C\#,
embodying a recursive session where a sequence of integer is received, and then the sum of them is sent back:
\begin{lstlisting}[numbers=none,xleftmargin=8pt]
class SumSrv : Recv<int, Offer<SumSrv, Send<int, End>>> { ... }
\end{lstlisting}
Although it is possible to declare such {\em session types} like above,
what we need is a {\em duality witness} (proof) encoded in C\#.
Consider a duality relation defined as a class, stating that \lstinline!Recv<V, S>! is a dual of \lstinline!Send<V, T>! if \lstinline!S! is a dual of \lstinline!T!:
\begin{lstlisting}[numbers=none,xleftmargin=8pt]
class DualRecv<V, Cont> : Dual<Recv<V, ...>, Dual<Send<V, ...>>> { ... }
\end{lstlisting}
We must refer to the two components of \lstinline!Cont! in the two ellipsis parts \lstinline!...!,
which would look like the following pseudo-code:
\begin{lstlisting}[numbers=none,xleftmargin=8pt]
// pseudo-C# code!
class DualRecv<V, Cont> : Dual<Recv<V, Cont@@.S@@>, Dual<Send<V, Cont@@.T@@>>> { ... }
\end{lstlisting}
which is not possible in C\# for now. One might recall {\em traits} or {\em type members} in C++ and Scala,
and {\em associated types} in Haskell \cite{DBLP:conf/popl/ChakravartyKJM05}.
There exists an encoding from
C\#'s F-bounded polymorphism to {\em family polymorphism} \cite{DBLP:journals/jot/SaitoI08},
at the cost of much boilerplate code.
That said, the use of recursive generic types seems promising,
and we are currently seeking a better design for recursive protocol combinators.

\section{Application}\label{sec:application}

\begin{figure}[tb]
 \begin{lstlisting}[multicols=2]
var cliChs = prot.Parallel(ids, (srvCh, id) => {
 for (var loop = true; loop;) {^\label{line:bitcoin:srv:outerloop}^
  srvCh.Offer(left: cont => {^\label{line:bitcoin:srv:offer}^
   var srvCh2 = cont.Receive(out var block)^\label{line:bitcoin:srv:receive}^
                    .DelegRecv(out var stopCh);^\label{line:bitcoin:srv:delegrecv}^
   stopCh.ReceiveAsync(out Task stop).Close();^\label{line:bitcoin:srv:receiveasync}^
   var miner = new Miner(block, id);^\label{line:bitcoin:srv:newminer}^
   while (true) {
    if (miner.TestNextNonce(out var nonce)) {^\label{line:bitcoin:srv:iffound}^
     srvCh = srvCh2.SelectLeft()^\label{line:bitcoin:srv:found}^
                   .Send(nonce).Goto0();
     break; // back to Offer() again^\label{line:bitcoin:srv:foundbreak}^
    } else if (stop.IsCompleted) {^\label{line:bitcoin:srv:ifstop}^
     srvCh = srvCh2.SelectRight().Goto0();
     break; // back to Offer() again^\label{line:bitcoin:srv:stopbreak}^
    } else { continue; }}},^\label{line:bitcoin:srv:continue}^
  right: end =>
   { end.Close(); loop = false; });}});^\label{line:bitcoin:srv:terminate}^
  \end{lstlisting}
  \caption{A Bitcoin Miner Server\label{fig:bitcoin:server}}
\end{figure}

As a more interesting application of \ourlibrary, we show a {\em Bitcoin miner},
where a collection of threads {\em iteratively} try
to find a {\em nonce} of the specified {\em block}.
The protocol for the Bitcoin miner is the following:
\begin{lstlisting}[numbers=none,xleftmargin=8pt]
var prot = Select(left: Send(Val<Block>, Deleg(chan:Recv(Unit,End),
                          Offer(left:Recv(Val<uint>,Goto0), right:Goto0))),
                  right: End);
\end{lstlisting}
The endpoint implementation is in \Cref{fig:bitcoin:server}.
The \lstinline!Parallel! method runs multiple threads in parallel,
by passing the anonymous function a pair of
the server channel \lstinline!srvCh! and an extra argument \lstinline!id!
which is extracted from the array of parameters \lstinline!ids!.
The client asks (\lstinline!Select!) a server thread to start the calculation by selecting \lstinline!left! label,
and then it sends a bitcoin \lstinline!Block! and a cancellation channel in a row.
Dually, after the server enters the
main loop in Line~\ref{line:bitcoin:srv:outerloop},
it offers a binary choice in Line~\ref{line:bitcoin:srv:offer},
receives the block and a channel in Line~\ref{line:bitcoin:srv:receive}-\ref{line:bitcoin:srv:delegrecv},
and then schedules asynchronous reception of cancellation in Line~\ref{line:bitcoin:srv:receiveasync}.
After that, the server starts the calculation
in Lines~\ref{line:bitcoin:srv:newminer}-\ref{line:bitcoin:srv:iffound},
entering the loop.
Meanwhile, the client waits for the server (\lstinline!Offer!),
and if it sees \lstinline!left! label,
then it receives a nonce of an unsigned integer (\lstinline!uint!).
The corresponding behaviour in the server is found in
Lines~\ref{line:bitcoin:srv:found}-\ref{line:bitcoin:srv:foundbreak},
where the server goes back to Line~\ref{line:bitcoin:srv:offer}.
In case another thread finds the nonce,
the client asynchronously sends cancellation to the server,
which is observed by the server in Lines~\ref{line:bitcoin:srv:ifstop}-\ref{line:bitcoin:srv:stopbreak},
notifying the \lstinline!right! label back to the client.
In the both case, the client returns to the beginning (\lstinline!Goto0!).
If nonce is not found and cancellation is not asked,
in Line~\ref{line:bitcoin:srv:continue},
the server tries the next iteration
without interacting with the client.
By selecting \lstinline!right! label at the top, the client can ask the server to terminate,
where the server closes the session and
assigns \lstinline!false! to \lstinline!loop! variable
in Line~\ref{line:bitcoin:srv:terminate}, exiting the outer loop.

\section{Concluding Remarks}\label{sec:conclusion}

We proposed \ourlibrary, a session-typed communication library for C\#.
The mainstream languages like C\# has not been targeted as
a platform implementing session-typed library,
where one of the reasons is that the type system of the language is
not suitable to implement them ---
they are less capable than
other languages like Haskell, Scala, F\# and OCaml,
in the sense of having richer type inference or
type-classes or implicits.
Another reason would be that the type system of C\# is considered quite similar to Java's one.
We proclaim that the {\em language features} like \lstinline!out!
variables (and closures) also matters for
establishing a safe, usable session communication pattern on top of it,
including session intervention,
as we have shown in the several examples in this paper.

The typestate approach taken by
StMungo by Kouzapas \etal \cite{DBLP:conf/ppdp/KouzapasDPG16}
equips session types on top of
programming front-end Mungo.
Gerbo and Padovani \cite{DBLP:conf/ppdp/KouzapasDPG16} also implements
session types in typestate-based encoding
via code generation using Java's annotation, 
enabling concurrent type-state programming a concise manner,
at the cost that the protocol conformance is checked dynamically.
On the other hand, type-states are sometimes {\em manually}
maintained via variable (re)assignment in \ourlibrary,
which weakens the static conformance checking.
However, we hope that sticking to the library-based implementation
with dynamic linearity checking
competes to the aforementioned tools
by providing the idiomatic usage of fluent interface.

The techniques and patterns incorporated in improved fluent interface in \ourlibrary is orthogonal
to tool support,
and we see opportunities to build them in combination
with other proposals like Scribble,
resulting in a concise multiparty session programming environment.
Notably, we see that the session intervention pattern is
also effective in multiparty setting.
We observe several instances of the fluent interface in Scribble family,
albeit without \lstinline!out! parameters,
in Java \cite{scribble,hy16,fase17}, Scala \cite{scalas17linear}, Go \cite{CHJNY2019}, and F\# \cite{NeykovaHYA18},
providing {\em multiparty session types}.
Code completion shown in the Introduction is also available
in various implementation in Scribble,
and most notably, the work by Neykova \etal \cite{NeykovaHYA18}
integrates Scribble with {\em Type Provider} in F\#.
SJ by Hu \etal \cite{sessionjava} extends Java with session primitives,
and also studies the protocol for session delegation in a distributed setting.


The protocol combinators are highly inspired from
Pucella and Tov's encoding of duality \cite{haskell08}.
To the author's knowledge,
the addition of delegation and recursion to \cite{haskell08} is new.
We believe the simplification of recursion
adds more readability to programs using protocol-combinator based implementations.
Scalas \etal \cite{scalas16lightweight}
and Padovani \cite{padovani16simple}
implements binary session types based
on duality encoded in linear i/o types by Dardha \etal \cite{DBLP:journals/iandc/DardhaGS17}.
While it does not require any intermediate object like protocol combinators,
we see the encoded session types sometimes have type readability issue,
as it makes a nested, flipping sequence inside i/o type constructors,
as mentioned in \cite[\S~6.2]{imai18sessionscp}.
Imai \etal \cite{imai18sessionscp} solved this readability issues via
{\em polarised session types}, at the cost of having polarity in types.
Albeit the lack of session type inference in \ourlibrary,
we also see the {\em explicit} approach taken by protocol combinators is
not a big obstacle, as it is also the case in C\# to declare method signatures explicitly.




\paragraph{Acknowledgements}
We thank the reviewers for thorough review and helpful comments for improving this paper.
This work is partially supported by KAKENHI 17K12662 from JSPS, Japan,
and by
Grants-in-aid for Promotion of Regional Industry-University-Government Collaboration from Cabinet Office, Japan.

\bibliographystyle{eptcs}
\bibliography{session}

\appendix
\section{Protocol Combinators}\label{app:combinators}
The protocol combinators are implemented as the following C\# static methods and fields.
The ellipsis parts are obvious return statements like ``\lstinline!return new Dual<Send<V,S>,Recv<V,T>>()!'':

\begin{adjustbox}{width=\columnwidth,center}
\begin{lstlisting}[numbers=none]
class ProtocolCombinators {
  static Val<V> Val<V>() { return new Val<V>(); }
  static Dual<Eps,Eps> End = new Dual<Eps,Eps>;    static Dual<Goto0,Goto0> Goto0 = new Dual<Goto0,Goto0>;
  static Dual<Send<V,T>,Recv<V,T>>  Send<V,T> (Func<Val<V>> v, Dual<V,T> cont) {...}
  static Dual<Recv<V,T>,Send<V,T>>  Recv<V,T> (Func<Val<V>> v, Dual<V,T> cont) {...}
  static Dual<Select<SL,SR>,Offer<TL,TR>>  Select<SL,SR,T0,T1> (Dual<SL,TL> contL, Dual<SR,TR> contR) {...}
  static Dual<Offer<SL,SR>,Select<TL,TR>>  Offer<SL,SR,T0,T1> (Dual<SL,TL> contL, Dual<SR,TR> contR) {...}
  static Dual<Deleg<S0,T0,S>,DelegRecv<S0,T>>  Deleg<S0,T0,S,T> (Dual<S0,T0> deleg, Dual<S,T> cont) {...}
  static Dual<DelegRecv<S0,S>,Deleg<S0,T0,T>>  DelegRecv<S0,T0,S,T> (Dual<S0,T0> deleg, Dual<S,T> cont) {...}
}
\end{lstlisting}
\end{adjustbox}

Modifiers such as \lstinline!public! are omitted.
Also, there is a small hack: The use of \protect{\lstinline!Func!} in the payload of \protect{\lstinline!Send!} and \protect{\lstinline!Recv!}
enables omitting the parenthesis \lstinline!()! after \lstinline!Val<V>! in protocol specifications  (as in \lstinline!prot! in \S~\ref{sec:introduction}).

\section{More on Recursion and Delegation}\label{app:delegation}
Delegation is also extended to handle with mutual recursive sessions:
\begin{lstlisting}[numbers=none,xleftmargin=8pt]
  static Session<S,E> DelegRecv<S1,S2,S,E>(this Session<DelegRecv<(S1,S2),S>,E> ch,
                                           out Session<S1,(S1,S2)> ch2) { ... }
  static Session<S,E> Deleg<S1,T1,S2,T2,S,E>(this Session<Deleg<(S1,S2),(T1,T2),S>,E> ch,
                                             Session<S1,(S1,S2)> ch2) { ... }
  static Session<S,E> DelegNew<S1,T1,S2,T2,S,E>(this Session<Deleg<(S1,S2),(T1,T2),S>,E> ch,
                                                out Session<T1,(T1,T2)> ch2) { ... }
\end{lstlisting}
To cope with the delegation in the {\em middle} of the session,
we further extend the communication API for delegation, as follows:
\begin{lstlisting}[numbers=none,xleftmargin=8pt]
  static Session<S,E> DelegRecv<S0,S1,S,E>(this Session<DelegRecv<(S0,S1),S>,E> ch,
                                           out Session<S0,S1> ch2) { ... }
  static Session<S,E> Deleg<S0,T0,S1,T1,S,E>(this Session<Deleg<(S0,S1),(T0,T1),S>,E> ch,
                                             Session<S0,S1> ch2) { ... }
  static Session<S,E> DelegNew<S0,T0,S1,T1,S,E>(this Session<Deleg<(S0,S1),(T0,T1),S>,E> ch,
                                                out Session<T0,T1> ch2) { ... }
\end{lstlisting}
It enables a session delegated in the middle of it
by having different session types in a session environment,
as in \lstinline!Session<S0,S1>! above.

\section{More Examples}
\label{app:moreexamples}
\Cref{fig:travelagency} is an implementation of a Travel Agency from \cite{sessionjava},
which incorporates two sessions in a distributed setting.
The {\em canceller} in Line~\ref{canceller} declared \lstinline!using!
modifier
stops the {\em registered} sessions in Lines~\ref{canceller1} and \ref{canceller2}
when scoping out,
which enables to
propagate connection failure in one of underlying TCP connections to the other.

We leave a few more examples for curious readers.
\Cref{fig:parallelhttp} is a {\em parallel http downloader}
from \cite{CHJNY2019}, which utilises \lstinline!Parallel!
method defined on the protocol specification object.
\Cref{fig:bitcoinclient} is a client to Bitcoin miner shown in \S~\ref{sec:application}.
\Cref{fig:polygonclipping} is
an implementation of {\em parallel polygon clipping}
from \cite{haskell08}, where
\lstinline!Pipeline! creates a series of threads
connected by two session-typed channels of which session type is described in a protocol specification.

\begin{figure}
\begin{lstlisting}[multicols=2]
protCA.Listen(IPAddress.Any, 8888, srvCh => {
 using var c = new SessionCanceller();^\label{canceller}^
 c.Register(srvCh);^\label{canceller1}^
 for (var loop = true; loop;) {
  srvCh.Offer(srvQuot =>
   quote.Receive(out var dest).Send(90.00m)
        .Offer(srvAcpt => {
    var cliCh = protAS.Connect("1.1.1.1", 9999);
    c.Register(cliCh);^\label{canceller2}^
    cliCh.Send(dest)
         .Receive(out var date).Close();
    srvAcpt.Send(date).Close();
    loop = false;
   }, srvReject => {
    srvCh = srvReject.Goto();
   }), srvQuit => {
    srvQuit.Close();
    loop = false; });}});
\end{lstlisting}
\caption{A Travel Agency (Agency Part)\label{fig:travelagency}}
\end{figure}

\begin{figure}[tb]
 \begin{lstlisting}[multicols=2]
// Client implementation (main thread)
foreach (var block in Block.GetSampleBlocks()) {
  // Send a block to each thread
  var ch2s = ch1s.Map(ch1 =>
    ch1.SelectLeft().Send(block));
  //  external choice
  var (ch3s, cancelChs) = ch2s.Map(ch2 => {
    var offer = ch2.DelegRecv(out var cancalCh)
      .OfferAsync(some => {
        var _ch3 = some.Receive(out var nonce);
        return (_ch3.Goto(), nonce);
      }, none => {
        var ch3 = none.Goto();
        return (ch3, default(uint?));
      });
    return (offer, cancalCh);
  }).Unzip();
  // Wait for any single thread to respond
  await Task.WhenAny(ch3s);
  // Send cancellation to each thread
  cancelChs.ForEach(ch => ch.Send().Close());
  // Get channels and results from future object
  var (ch4s, results) =
    ch3s.Select(c => c.Result).Unzip();
  // Print results (omitted)
  // Assign and recurse
  ch1s = ch4s;
}
// No blocks to mine, finish channels
ch1s.ForEach(ch1 => ch1.SelectRight().Close());
\end{lstlisting}
\caption{A Bitcoin miner client\label{fig:bitcoinclient}}
\end{figure}

\begin{figure}[tb]
\begin{lstlisting}[multicols=2]
using System;
using System.Linq;
using System.Net.Http;
using System.Threading.Tasks;
using Session;
using Session.Threading;
using static ProtocolCombinator;

public class Program {
  public static async Task Main(string[] args) {
    // Protocol specification
    var prot = Select(left: Send(Val<string>,
      Recv(Val<byte[]?>, Goto0)), right: End);

    var n = Environment.ProcessorCount;
    var ch1s = prot.Parallel(n, ch1 => {
      // Init http client
      var http = new HttpClient();

      // Work...
      for (var loop = true; loop;) {
        ch1.Offer(left => {
          var ch2 = left.Receive(out var url);
          var data = Download(url);
          ch1 = ch2.Send(data).Goto();
        }, right => {
          right.Close();
          loop = false;
        });
      }

      // Download function
      byte[]? Download(string url) {
        try {
          return http
            .GetByteArrayAsync(url).Result;
        } catch {
          return null;
        }
      }
    });

    // Pass jobs to each thread
    var (ch2s, ch1s_rest, args_rest) =
      ch1s.ZipWith(args, (ch1, arg) => {
        var ch3 = ch1.SelectLeft().Send(arg)
          .ReceiveAsync(out var data);
        return (ch3.Sync(), data);
      });

    // Close unneeded channels
    ch1s_rest
      .ForEach(c => c.SelectRight().Close());

    var (working, results) = ch2s.Unzip();
    var working_list = working.ToList();
    var result_list = results.ToList();

    // Wait for a single worker finish
    // and pass a new job
    foreach (var url in args_rest) {
      var finished =
        await Task.WhenAny(working_list);
      working_list.Remove(finished);
      var ch3 = (await finished).Goto()
        .SelectLeft().Send(url)
        .ReceiveAsync(out var data);
      working_list.Add(ch3.Sync());
      result_list.Add(data);
    }

    // Wait for still working threads
    while(working_list.Any())
    {
      var finished =
        await Task.WhenAny(working_list);
      working_list.Remove(finished);
      (await finished).Goto()
                      .SelectRight().Close();
    }

    // Save to files or something...
  }
}

\end{lstlisting}
\caption{Parallel HTTP Downloader \cite{CHJNY2019}\label{fig:parallelhttp}}
\end{figure}

\begin{figure}[tb]
\begin{lstlisting}[multicols=2]
using System;
using System.Collections.Generic;
using Session;
using Session.Threading;
using static ProtocolCombinator;

public class Program {
  public static void Main(string[] args) {
    // Input: clippee
    var vertices = new Vector[] {
      new Vector(2.0, 2.0),
      new Vector(2.0, 6.0),
      // ... and more points
    };

    // Input: clipper
    var clipper = new Vector[]
    {
      new Vector(1.0, 3.0),
      new Vector(3.0, 6.0),
      // ... and more points
    };

    // Split clipper each edges
    var edges =
      new (Vector, Vector)[clipper.Length];
    for (int i = 0; i < edges.Length; i++) {
      edges[i] = (clipper[i],
        clipper[(i + 1) % clipper.Length]);
    }

    // Protocol specification
    var prot = Select(left: Send(Val<Vector>,
      Goto0), right: End);

    var (in_ch, out_ch) = prot.Pipeline(edges,
      // Each thread
      (prev1, next1, edge) => {
        Vector? first = null;
        Vector from = default;
        Vector to = default;
        for (var loop = true; loop;) {
          prev1.Offer(left => {
            var prev2 = left
              .Receive(out var vertex);
            from = to;
            to = vertex;
            if (first == null) {
              first = to;
            } else {
              var clipped =
                Clip((from, to), edge);
              foreach (var v in clipped) {
                next1 = next1
                  .SelectLeft().Send(v).Goto();
              }
            }
            prev1 = prev2.Goto();
          }, right => {
            var clipped =
              Clip((to, first.Value), edge);
            foreach (var v in clipped) {
              next1 = next1.SelectLeft()
                           .Send(v).Goto();
            }
            next1.SelectRight();
            loop = false;
          });
        }
      }
    );

    // Main thread
    // Send vertices to pipeline
    foreach (var v in vertices) {
      in_ch = in_ch.SelectLeft().Send(v).Goto();
    }
    in_ch.SelectRight().Close();

    // Collect result from pipeline
    var result = new List<Vector>();
    for (var loop = true; loop;) {
      out_ch.Offer(left => {
        out_ch = left.Receive(out var vertex)
                     .Goto();
        result.Add(vertex);
      }, right => {
        right.Close();
        loop = false;
      });
    }

    // Print result
    for (int i = 0; i < result.Count; i++) {
      Console.WriteLine(result[i]);
    }
  }
}
\end{lstlisting}
\caption{Polygon Clipping Pipeline \cite{haskell08}\label{fig:polygonclipping}}
\end{figure}


\end{document}